\documentclass[showpacs,amsmath,
twocolumn,
aps,prl]{revtex4}
\usepackage{graphicx}
\usepackage{dcolumn}

\begin{document}

\title{Current induced spin polarization in strained semiconductors}
\author{Y. Kato}
\author{R. C. Myers}
\author{A. C. Gossard}
\author{D. D. Awschalom}
\affiliation{Center for Spintronics and Quantum Computation, 
University of California, Santa Barbara, CA 93106}
\date{\today}
\begin{abstract}
The polarization of conduction electron spins due to an electrical current 
is observed in strained nonmagnetic semiconductors using static and 
time-resolved Faraday rotation. The density, lifetime, and orientation rate 
of the electrically-polarized spins are characterized by a combination of 
optical and electrical methods. In addition, the dynamics of the 
current-induced spins are investigated by utilizing electrical pulses 
generated from a photoconductive switch. These results demonstrate the 
possibility of a spin source for semiconductor spintronic devices without 
the use of magnetic materials.
\end{abstract}
\maketitle
Preparation and control of spin information are key issues in the 
development of spintronics \cite{Wolf:2001,Semiconductor:2002}. 
The use of nonmagnetic semiconductors to electrically control electron spins 
has been demonstrated \cite{Salis:2001,Kato:2003}, even in the 
absence of magnetic fields \cite{Kato:2004}. An electrical means of 
preparing spin-polarized carriers without magnetic materials would provide a 
further step toward all-electrical nonmagnetic spintronic devices. It has 
been proposed that current-induced spin polarization may provide such an 
opportunity. The existence of a spin current perpendicular to a charge 
current, which would cause spin accumulation at the edges of a sample, has 
been predicted 
\cite{Dyakonov:1971,Hirsch:1999,Zhang:2000,Murakami:2003}. 
There also exist theories for spatially homogeneous spin polarization 
resulting from an electrical current in systems such as two dimensional 
electron gases 
\cite{Levitov:1985,Edelstein:1990,Aronov:1991,Magarill:2001,Chaplik:2002,%
Inoue:2003,Cartoixa:2001,Silsbee:2001}. 
It is anticipated that application of an electric field establishes an 
effective magnetic field which polarizes the electron spins. Experimental 
attempts to detect such polarization using ferromagnetic contacts have been 
made \cite{Hammer:1999,Hammer:2000,Hammer:2001}, but the 
local Hall effect and anisotropic magnetoresistance complicate these 
measurements 
\cite{Monzon:1999,van:1999,Hammer:2002,Monzon:2000,Filip:2000}.

Recently, it was found that strained semiconductors exhibit spin splitting 
in the presence of applied electric fields \cite{Kato:2004}. In this 
Letter, we report the optical detection of current-induced electron spin 
polarization in strained GaAs and InGaAs epitaxial layers. The high 
sensitivity of the optical Faraday rotation technique allows detection of 
100 spins in an integration time of about a second, unambiguously revealing 
the presence of a small spin polarization due to laterally applied electric 
fields. We are able to extract quantitative values of spin density by 
comparing the Faraday rotation due to electric fields to that induced by 
optical spin injection. The spin orientation process persists up to a 
temperature $T = 150$~K with no marked degradation of efficiency, and is 
also observed over picosecond timescales in time-resolved measurements in 
which a coherent spin population is excited with electrical pulses derived 
from a photoconductive switch.

The samples studied here are grown on (001) semi-insulating GaAs substrates 
by molecular beam epitaxy. Eight different heterostructures with strained 
In$_{0.07}$Ga$_{0.93}$As layers as well as strained GaAs membranes 
\cite{Kato:2004} were investigated. Qualitatively similar behavior 
has been seen in all samples, but for most of this Letter we will 
concentrate on devices fabricated from one of the wafers (sample E in 
Ref.~\cite{Kato:2004}) in order to avoid confusion. The 
heterostructure consists of 500~nm of $n$-In$_{0.07}$Ga$_{0.93}$As (Si-doped 
for $n = 3\times 10^{16}$~cm$^{ - 3})$ capped with 100~nm of undoped GaAs. 
The $n$-InGaAs layer is strained due to the lattice mismatch 
\cite{Jain:1996}, and although strain relaxation can be seen, there 
exists residual strain as determined by reciprocal space mapping with an 
x-ray diffractometer at room temperature. 

\begin{figure}[b]\includegraphics{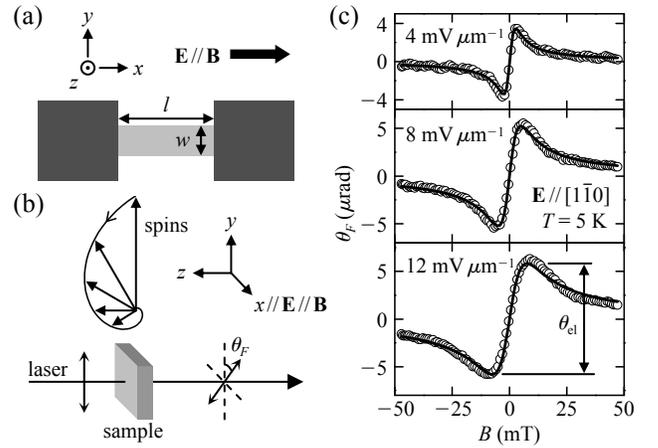}\caption{\label{fig1}
(a) Device schematic and sample geometry. Dark areas are contacts and light 
gray area is the InGaAs channel. (b) Schematic of experimental setup and 
Larmor precession of spins excited by electrical current. (c) 
Voltage-induced $\theta _{F}$ as a function of $B$ for $E = 4$, 8, and 
12~mV~$\mu $m$^{ - 1}$ (\textbf{E}//$[1\bar {1}0])$, taken at $T = 5$~K from 
a device with $w = 60$~$\mu $m and $l = 200$~$\mu $m. Open circles are data 
and lines are fits as explained in the text. Constant offsets have been 
subtracted for clarity.
}\end{figure}
A schematic of a device is shown in Fig.~\ref{fig1}(a). Photolithography and chemical 
wet etching are employed to form the $n$-InGaAs mesa, and Ni/GeAu metallization 
followed by annealing is used to make ohmic contacts to the $n$-InGaAs layer. 
Two such devices are fabricated on a chip to allow measurements with the 
electric field \textbf{E} along either of the two crystal directions [110] 
and $[1\bar {1}0]$. The sample is placed inside a magneto-optical cryostat 
with a variable temperature insert such that the magnetic field \textbf{B} 
is parallel to \textbf{E}. In order to probe the spin polarization in the 
sample, Faraday rotation is measured in the Voigt geometry [Fig.~\ref{fig1}(b)]. A 
mode-locked Ti:sapphire laser operating at a repetition frequency 
$f_{\text{rep}} = 75.5$~MHz produces $\sim $150~fs pulses and is tuned to a 
wavelength $\lambda = 867$~nm. A linearly polarized probe beam with an 
average power of 130~$\mu $W is directed along the $z$ axis, normally incident 
and focused on the sample. The polarization axis of the transmitted beam 
rotates by an angle that is proportional to the $z$ component of the spins 
\cite{Crooker:1997}, and the rotation angle $\theta _{F}$ is detected 
with a balanced photodiode bridge. A square-wave voltage with peak-to-peak 
value $V_{\text{pp}} $ at frequency $f_1 = 51.2$~kHz is applied to one of 
the contacts while the other contact is grounded. An alternating electric 
field with amplitude $E = V_{\text{pp}} / \left( {2l} \right)$ is 
established along the InGaAs channel of width $w$ and length $l$, assuming 
negligible contact resistance. The current-induced $\theta _{F}$ is 
lock-in detected at $f_{1}$ as a function of the applied magnetic field $B$ 
along the $x$ axis. Typical data are shown in Fig.~\ref{fig1}(c), for three different 
electric fields. 

The curves can be explained by assuming a constant orientation rate for 
spins polarized along the $y$ axis. In a manner similar to the case of optical 
orientation \cite{Optical:1984} under static illumination, the $z$ component 
of spin per unit volume $\rho _{z}$ can be written as
\[
\rho _z = \int_0^\infty {dt\left[ {\gamma \exp \left( { - t / \tau } 
\right)\sin \left( {\omega _L t} \right)} \right]} = \frac{\rho _{\text{el}} 
\omega _L \tau }{\left( {\omega _L \tau } \right)^2 + 1},
\]
where $\gamma $ is the number of spins oriented along the $y$ axis per unit 
time per unit volume, $\tau $ is the inhomogeneous transverse spin lifetime, 
$\omega _L = g\mu _B B / \hbar $ is the electron Larmor frequency, $\rho 
_{\text{el}} \equiv \gamma \tau $ is the steady-state spin density due to 
electrical excitation, $g$ is the effective electron g factor, $\mu _{B}$ is 
the Bohr magneton, and $\hbar $ is the Planck constant. The upper 
integration limit is taken as infinity since the modulation period is much 
longer than $\tau $, and the effective magnetic field observed in 
\cite{Kato:2004} is omitted as it is a second order effect due to its 
parallel orientation to the current-induced spins. Approximating the beam 
profiles as Gaussians, assuming spatially uniform $\rho _{\text{el}} $, and 
letting $\theta _{F}$ be proportional to both $\rho _{z}$ and the 
intensity of the probe beam with a proportionality constant $A$, we find 
\begin{eqnarray*}
 \theta _F\! &=& Ad\int\!\!\!\int {dxdy\left[ {\frac{\rho _{\text{el}} \omega _L 
\tau }{\left( {\omega _L \tau } \right)^2 + 1}I_p \exp \left( { - 
\frac{x^2}{2\sigma _x ^2} - \frac{y^2}{2\sigma _y ^2}} \right)} \right]} \\ 
 &=& 2\pi Ad\rho _{\text{el}} I_p \sigma _x \sigma _y \frac{\omega _L \tau 
}{\left( {\omega _L \tau } \right)^2 + 1} = \theta _{\text{el}} \frac{\omega 
_L \tau }{\left( {\omega _L \tau } \right)^2 + 1}, 
\end{eqnarray*}
where $d$ is the thickness of the epitaxial film, $I_{p}$ is the peak intensity 
of the probe beam, $\sigma _{x}$ and $\sigma _{y}$ are the standard 
deviation of the laser intensity in the $x$ and $y$ directions, respectively, and 
$\theta _{\text{el}} \equiv 2\pi Ad\rho _{\text{el}} I_p \sigma _x \sigma _y 
$ is the amplitude of the electrically-induced $\theta _{F}$. By fitting 
the data with the above equation, we obtain $\theta _{\text{el}} $ and 
$\omega _{L}\tau $. The data does not change significantly when the 
laser spot is moved across or along the channel, while polarization due to a 
spin current is expected to appear only within a spin diffusion length from 
the channel edge and reverse sign for opposing edges 
\cite{Dyakonov:1971,Hirsch:1999,Zhang:2000,Murakami:2003}. 
This suggests that the relevant mechanism is wave-vector dependent 
spin-splitting which predicts uniform polarization 
\cite{Edelstein:1990,Aronov:1991,Magarill:2001,Chaplik:2002,Inoue:2003,%
Cartoixa:2001,Silsbee:2001}.

\begin{figure}\includegraphics{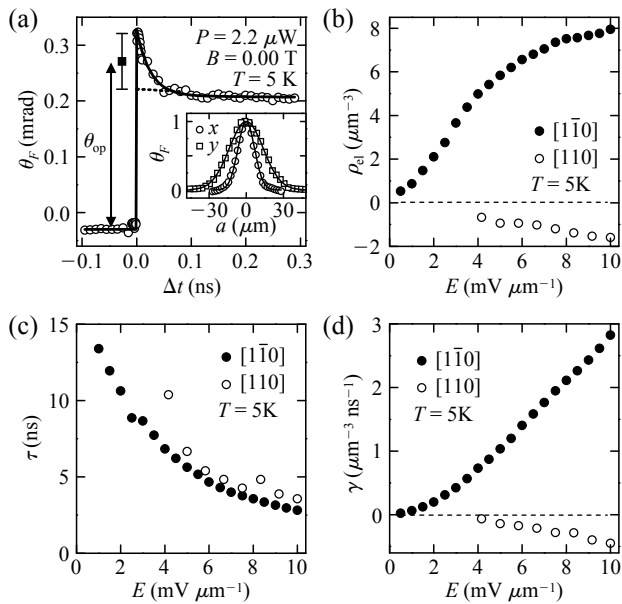}\caption{\label{fig2}
(a) Time-resolved Faraday rotation data in the absence of electric fields at 
$T = 5$~K and $B = 0.00$~T showing $\theta _{\text{op}} $ (filled square) 
due to circularly polarized pump pulse with $P = 2.2$~$\mu $W. Open circles 
are data, the dotted line shows the extrapolation used for obtaining $\theta 
_{\text{op}} $, and the solid line is a guide to the eye. Inset shows the 
spatial scans of $\theta _{F}$ taken in the absence of electric field at 
$T = 5$~K and $B = 0.00$~T. $\Delta t = 50$~ps for the scan along $x$ (open 
circles) and $\Delta t = 10$~ps for the scan along $y$ (open squares). Lines 
are Gaussian fits used to extract beam profiles. (b), (c), and (d) show the 
spin density $\rho _{\text{el}} $, the spin lifetime $\tau $, and the spin 
orientation rate $\gamma $, respectively, as a function of $E$ for 
\textbf{E}//$[1\bar {1}0]$ (filled circles, taken from a device with $w = 
60$~$\mu $m and $l = 200$~$\mu $m) and \textbf{E}//[110] (open circles, 
taken from a device with $w = 80$~$\mu $m and $l = 300$~$\mu $m). Overall 
scaling of $\rho _{\text{el}} $ and $\gamma $ has errors up to 40{\%}.
}\end{figure}
Additional quantitative analysis can be performed by measuring time-resolved 
Faraday rotation (TRFR) \cite{Crooker:1997} in the absence of electric 
fields with the same probe power. In this measurement, a circularly 
polarized pump beam (13~$\mu $W, $\lambda = 867$~nm) is focused onto an 
overlapping spot with the probe beam, optically injecting electron spins 
\cite{Optical:1984}. The time delay $\Delta t$ between the pump pulse and 
the probe pulse is adjusted by a mechanical delay line, and the pump beam 
helicity is modulated at 50.1~kHz with a photoelastic modulator for lock-in 
detection. In the presence of an applied magnetic field, $\theta _{F}$ 
oscillates as a function of $\Delta t$ with frequency $\omega _{L}$, from 
which $g$ can be determined \cite{Crooker:1997}. Measurement at $B = 0.5$~T 
gives $\left| g \right| = 0.64$, which is used to obtain $\tau $ from the 
fits to the voltage-induced Faraday rotation data. Furthermore, $\rho 
_{\text{el}} $ can be extracted by comparing $\theta _{\text{el}} $ to the 
optically induced $\theta _{F}$. Just after the pump pulse, the spin 
density profile should track the pump beam intensity profile, and the 
Faraday rotation is
\begin{eqnarray*}
 \theta _F &=& Ad\int\!\!\!\int {dxdy\left[ {\rho _{\text{op}} I_p \exp \left( 
{ - \frac{x^2}{\sigma _x ^2} - \frac{y^2}{\sigma _y ^2}} \right)} \right]} 
\\ 
 &=& \pi Ad\rho _{\text{op}} I_p \sigma _x \sigma _y \equiv \theta 
_{\text{op}} ,  
\end{eqnarray*}
\begin{table}[b]
\caption{\label{table1}
Summary of the strain-induced spin-splitting coefficient $\beta $ and spin 
orientation efficiency $\eta $. All data taken at $T = 5$~K from devices 
with $w = 80$~$\mu $m and $l = 300$~$\mu $m. The same device was used for 
the measurement of $\beta $ and $\eta $ for a given wafer.
}
\begin{ruledtabular}
\begin{tabular}
{@{}c@{}cr@{}rcr@{}rc@{}c@{}r@{}r@{}c@{}r@{}r@{}r@{}r@{}ccr@{}r@{}c@{}r@{}r@{}r@{}r@{}c@{}}
& 
\multicolumn{7}{c@{}}{$\beta $ (neV ns $\mu $m$^{ - 1})\footnotemark[1]$}  &&
\multicolumn{17}{c}{$\eta $ ($\mu $m$^{ - 2}$ V$^{ - 1}$ ns$^{ - 1})$}  \\
Sample\footnotemark[1]&& 
\multicolumn{2}{c}{[110]} && 
\multicolumn{2}{c}{$[1\bar {1}0]$} &&& 
\multicolumn{8}{c}{[110]} && 
\multicolumn{8}{c}{$[1\bar {1}0]$}  \\
\hline
GaAs&\hspace*{6pt}& \multicolumn{2}{c}{$b$} &\hspace*{6pt}& +& 99&\hspace*{6pt}&& 
\multicolumn{8}{c}{$b$} && +& 216&(&+& 92&$-$& 67&) \\
B&& $-$&  13&& $-$& 39&&& $-$& 35&(&+& 13&$-$& 22&)&& +&   4&(&+&   2&$-$&   1&) \\
C&&  + &  44&& $-$& 21&&& $-$& 15&(&+&  7&$-$& 26&)&& +&  65&(&+& 113&$-$&  32&) \\
D&&  + & 125&&  + &  4&&&  + &  4&(&+& 20&$-$&  2&)&& +&  30&(&+& 158&$-$&  17&) \\
E&&  + & 112&&  + & 27&&& $-$& 61&(&+& 19&$-$& 27&)&& +& 422&(&+& 185&$-$& 133&) \\
F&&  + &  84&&  + & 23&&&  + & 75&(&+& 19&$-$& 28&)&& +& 216&(&+& 143&$-$&  81&) \\
G&&  + &  75&&  + & 13&&&  + & 25&(&+& 20&$-$& 10&)&& +&  36&(&+&  28&$-$&  14&) \\
H&&  + &  23&& $-$& 26&&& $-$& 64&(&+& 23&$-$& 40&)&& +&  84&(&+&  52&$-$&  31&) \\
I&&  + &  89&&  + & 42&&&  + &  7&(&+&  4&$-$&  2&)&& +&  25&(&+&  14&$-$&   9&) \\
\end{tabular}
\end{ruledtabular}
\footnotetext[1]{Sample details and original data on $\beta $ given in reference \cite{Kato:2004}.}
\footnotetext[2]{The strained membrane sample only had a device along $[1\bar {1}0]$.}
\end{table}
where $\rho _{\text{op}} $ is the peak spin density due to the pump beam and 
$\theta _{\text{op}} $ is the optically induced $\theta _{F}$. In 
Fig.~\ref{fig2}(a), TRFR data at early time is shown. Initial rapid decay is 
attributed to the presence of holes and/or excitons \cite{Kikkawa:1997}, 
and since their contribution to $\theta _{F}$ is unknown, the value used 
for $\theta _{\text{op}} $ is an average of the maximum $\theta _{F}$ at 
$\Delta t = 0$ and the value extrapolated back to $\Delta t = 0$ from the 
data points after the rapid decay. These two values were also used to set 
the bounds on $\theta _{\text{op}} $. Assuming 50{\%} polarization from 
circularly polarized light \cite{Optical:1984}, the total number of 
optically injected spins per pulse is $2\pi \rho _{\text{op}} \sigma _x 
\sigma _y d = \left( {P\lambda } \right) / \left( {4\pi \hbar 
cf_{\text{rep}} } \right)$, where $P$ is the absorbed power of the pump beam 
and $c$ is the speed of light. The reflected and the transmitted power of the 
pump beam are measured on and off the InGaAs mesa in order to determine $P$, 
while $\sigma _{x}$ and $\sigma _{y}$ are obtained from measurement of 
$\theta _{F}$ as a function of pump-probe spatial distance $a$ [Fig.~\ref{fig2}(a) 
inset] using a stepper-motor-driven mirror \cite{Kato:2004}. Taking 
signal convolution into account, normalized data are fit to $\exp \left[ { - 
a^2 / \left( {4\sigma _a ^2} \right)} \right]$ to give $\sigma _x = 
4.7$~$\mu $m and $\sigma _y = 9.7$~$\mu $m. We estimate 20 {\%} error in the 
determination of $\rho _{\text{op}} = \left( {P\lambda } \right) / \left( 
{8\pi ^2\sigma _x \sigma _y d\hbar cf_{\text{rep}} } \right)$. Finally, 
$\rho _{\text{el}} = \rho _{\text{op}} \theta _{\text{el}} / \left( {2\theta 
_{\text{op}} } \right)$ and $\gamma = \rho _{\text{el}} / \tau $ are 
obtained. 

As expected, $\rho _{\text{el}} $ increases with $E$ [Fig.~\ref{fig2}(b)] and reaches 
8~$\mu $m$^{ - 3}$, corresponding to a spin polarization of $2.7\times 10^{ 
- 4}$, while $\tau $ decreases with $E$ [Fig.~\ref{fig2}(c)], leading to the saturation 
of $\rho _{\text{el}} $. An approximately linear relation between $\gamma $ 
and $E$ is seen [Fig.~\ref{fig2}(d)], and we define the spin orientation efficiency 
$\eta $ as the slope of a linear fit to $\gamma $ versus $E.$ The slight 
nonlinearity seen for small $E$ may be due to the current-voltage 
characteristics of the device not being perfectly linear. The sign of $\rho 
_{\text{el}} $ is determined from current-induced nuclear polarization in 
\textbf{E}$ \bot $\textbf{B}$ \bot z$ geometry, measured by Larmor 
magnetometry \cite{Salis:2002}. We note that $\eta $ is more positive 
when \textbf{E}//$[1\bar {1}0]$, consistent throughout the eight 
heterostructures investigated (Table~\ref{table1}). Although theories for 
two-dimensional electron gas predict a proportional relation between 
spin-splittings and $\eta $ 
\cite{Edelstein:1990,Aronov:1991,Magarill:2001,Chaplik:2002,Inoue:2003,Cartoixa:2001}, 
we observe sign contradictions in some cases. 

\begin{figure}\includegraphics{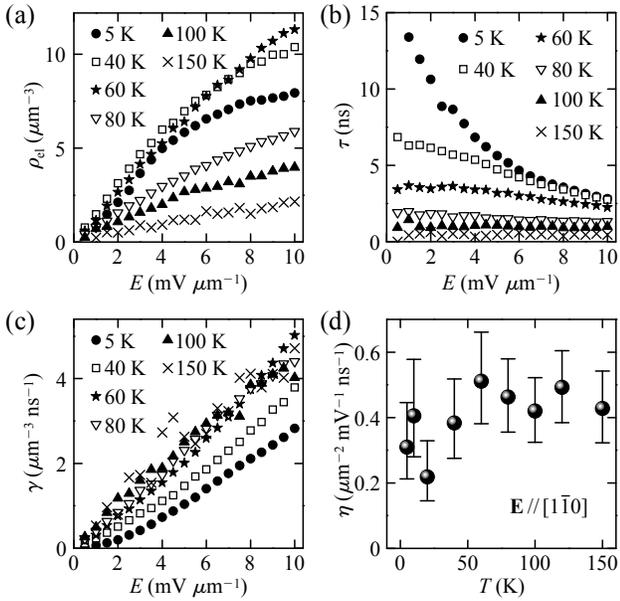}\caption{\label{fig3}
Temperature dependence of $\rho _{\text{el}} $ (a), $\tau $ (b), $\gamma $ 
(c), and $\eta $ (d). \textbf{E}//$[1\bar {1}0]$ and data taken from a 
device with $w = 60$~$\mu $m and $l = 200$~$\mu $m.
}\end{figure}

In Fig.~\ref{fig3}, we explore the temperature dependence of the effect. At each $T$, 
$\lambda $ is adjusted to optimize the Faraday rotation signal due to 
optically injected spins, and the set of measurements is repeated. At 
temperatures above 80 K, $\rho _{\text{el}} $ becomes smaller [Fig.~\ref{fig3}(a)] 
due to the decline of $\tau $ [Fig.~\ref{fig3}(b)]. It is seen that $\gamma $ and 
$\eta $ do not considerably change up to $T = 150$~K [Fig.~\ref{fig3}(c) and 
Fig.~\ref{fig3}(d)]. The signal is below the noise level for $T > 150$~K due to 
shorter $\tau $ and lower sensitivity of Faraday rotation. 

\begin{figure}\includegraphics{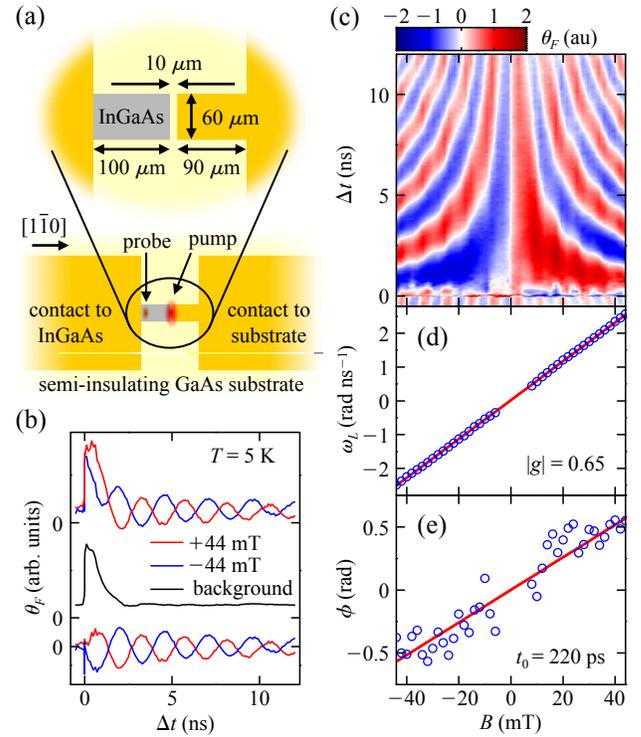}\caption{\label{fig4}
(a) Device schematic. (b) Time evolution of voltage induced $\theta _{F}$. 
Top curves (red, $B = + 44$~mT, blue, $B = - 44$~mT) show the raw data. 
Black curve is the background signal. Bottom curves show the data after 
background subtraction. (c) $\theta _{F}$ (background subtracted) as a 
function of $\Delta t$ and $B$. (d) and (e) show $\omega _{L}$ and $\phi $, 
respectively, obtained from fits to data in (c).
}\end{figure}

In order to investigate the effect at shorter timescales, we employ a 
two-color pump-probe setup \cite{Kikkawa:1997} in conjunction with a 
photoconductive switch \cite{Auston:1983} to produce electrical pulses 
[Fig.~\ref{fig4}(a)]. The pump beam (2.8~mW) is linearly polarized to avoid optical 
spin injection, and is tuned to $\lambda = 809$~nm in order to excite 
carriers in the GaAs substrate. The photoexcited carriers temporarily short 
the two contacts, thus producing an electrical pulse whose duration is 
limited by the carrier lifetime. The probe beam (130~$\mu $W, $\lambda = 
867$~nm) is placed $\sim $90 $\mu $m from the pump beam. A square-wave with 
$V_{\text{pp}} = 20$~V at $f_1 = 497$~Hz is applied on the contact to the 
InGaAs layer, and the contact to the substrate is grounded. The pump beam is 
chopped at $f_2 = 5.69$~kHz and the signal is lock-in detected at 
$f_{1}\pm  f_{2}$. The time evolution of voltage-induced $\theta _{F}$ at 
$B = \pm 44$~mT is shown in Fig.~\ref{fig4}(b), demonstrating current-induced 
electron spin polarization at these timescales. The sign of the signal 
reverses with the direction of $B$, as expected for in-plane excitation of 
spins. There exists a non-oscillating $B$-independent signal in addition to the 
oscillating component, which can be extracted by averaging the data over a 
range of $B$. This background may be an absorption signal arising from excess 
carriers, detected due to an imperfect balancing of the photodiode bridge. 
We subtract the background for analysis, and the data after subtraction are 
plotted at the bottom of Fig.~\ref{fig4}(b). Such data taken for a range of $B$ are 
plotted in Fig.~\ref{fig4}(c). The weak amplitude ripples along the $B$ axis are due to 
resonant spin amplification \cite{Kikkawa:1998}. We fit each $\theta 
_{F}$ versus $\Delta $t curve to $\theta _0 \exp \left( { - \Delta t / 
\tau } \right)\sin \left( {\omega _L \Delta t - \phi } \right)$, where 
$\theta _{0}$ is the initial amplitude and $\phi $ is the phase. We obtain 
$\left| g \right| = 0.65$ from the slope of $\omega _{L}$ [Fig.~\ref{fig4}(d)], 
consistent with measurements using optical spin injection, thus confirming 
that the voltage-induced signal arises from electron spins. There also 
exists a slope to $\phi $ [Fig.~\ref{fig4}(e)] due to the time delay $t_0 = \phi / 
\omega _L = 220$~ps between the pump pulse and the spin excitation, which we 
attribute to the width of the electrical pulse. 

The presented results demonstrate electron spin polarization due to an 
electrical current in strained semiconductors, generating possibilities for 
spintronic devices in which spin initialization and manipulation are 
performed without magnetic materials or magnetic fields. Larger spin 
polarization may be achievable in narrow gap semiconductors and 
two-dimensional quantum structures where the spin-orbit effects are larger. 

\begin{acknowledgments}
We acknowledge support from DARPA SPINS and QuIST programs, and the DMEA.
\end{acknowledgments}


\begin{thebibliography}{32}
\bibitem{Wolf:2001} S. A. Wolf \textit{et al.}, Science \textbf{294}, 1488 (2001).
\bibitem{Semiconductor:2002} \textit{Semiconductor Spintronics and Quantum Computation}, edited by D. D. Awschalom, D. Loss, and N. Samarth (Springer-Verlag, Berlin, 2002).
\bibitem{Salis:2001} G. Salis \textit{et al.}, Nature \textbf{414}, 619 (2001).
\bibitem{Kato:2003} Y. Kato \textit{et al.}, Science \textbf{299}, 1201 (2003).
\bibitem{Kato:2004} Y. Kato, R. C. Myers, A. C. Gossard, and D. D. Awschalom, Nature \textbf{427}, 50 (2004).
\bibitem{Dyakonov:1971} M. I. Dyakonov and V. I. Perel, Phys. Lett. A \textbf{35}, 459 (1971).
\bibitem{Hirsch:1999} J. E. Hirsch, Phys. Rev. Lett. \textbf{83}, 1834 (1999).
\bibitem{Zhang:2000} S. Zhang, Phys. Rev. Lett. \textbf{85}, 393 (2000).
\bibitem{Murakami:2003} S. Murakami, N. Nagaosa, and S. C. Zhang, Science \textbf{301}, 1348 (2003).
\bibitem{Levitov:1985} L. S. Levitov, Y. V. Nazarov, and G. M. Eliashberg, Sov. Phys. JETP \textbf{61}, 133 (1985).
\bibitem{Edelstein:1990} V. M. Edelstein, Solid State Commun. \textbf{73}, 233 (1990).
\bibitem{Aronov:1991} A. G. Aronov, Y. B. Lyanda-Geller, and G. E. Pikus, Sov. Phys. JETP \textbf{73}, 537 (1991).
\bibitem{Magarill:2001} L. I. Magarill, A. V. Chaplik, and M. V. Entin, Semiconductors \textbf{35}, 1081 (2001).
\bibitem{Chaplik:2002} A. V. Chaplik, M. V. Entin, and L. I. Magarill, Physica E \textbf{13}, 744 (2002).
\bibitem{Inoue:2003} J. Inoue, G. E. W. Bauer, and L. W. Molenkamp, Phys. Rev. B \textbf{67}, 33104 (2003).
\bibitem{Cartoixa:2001} X. Cartoixa, D. Z. Y. Ting, E. S. Daniel, and T. C. McGill, Superlattices Microstruct. \textbf{30}, 309 (2001).
\bibitem{Silsbee:2001} R. H. Silsbee, Phys. Rev. B \textbf{63}, 155305 (2001).
\bibitem{Hammer:1999} P. R. Hammer, B. R. Bennett, M. J. Yang, and M. Johnson, Phys. Rev. Lett. \textbf{83}, 203 (1999).
\bibitem{Hammer:2000} P. R. Hammer and M. Johnson, Phys. Rev. B \textbf{61}, 7207 (2000).
\bibitem{Hammer:2001} P. R. Hammer and M. Johnson, Appl. Phys. Lett. \textbf{79}, 2951 (2001).
\bibitem{Monzon:1999} F. G. Monzon, H. X. Tang, and M. L. Roukes, Phys. Rev. Lett. \textbf{84}, 5022 (1999).
\bibitem{van:1999} B. J. van Wees, Phys. Rev. Lett. \textbf{84}, 5023 (1999).
\bibitem{Hammer:2002} P. R. Hammer, B. R. Bennett, M. J. Yang, and M. Johnson, Phys. Rev. Lett. \textbf{84}, 5024 (1999).
\bibitem{Monzon:2000} F. G. Monzon and M. L. Roukes, J. Mag. Mag. Mat. \textbf{198-199,} 632 (1999).
\bibitem{Filip:2000} A. T. Filip \textit{et al.}, Phys. Rev. B \textbf{62}, 9996 (2000).
\bibitem{Jain:1996} S. C. Jain, M. Willander, and H. Maes, Semicond. Sci. Technol. \textbf{11}, 641 (1996).
\bibitem{Crooker:1997} S. A. Crooker \textit{et al.}, Phys. Rev. B \textbf{56}, 7574 (1997).
\bibitem{Optical:1984} \textit{Optical Orientation}, edited by F. Meier and B. P. Zakharchenya (Elsevier, Amsterdam, 1984).
\bibitem{Kikkawa:1997} J. M. Kikkawa, I. P. Smorchkova, N. Samarth, and D. D. Awschalom, Science \textbf{277}, 1284 (1997).
\bibitem{Salis:2002} G. Salis, D. D. Awschalom, Y. Ohno, and H. Ohno, Phys. Rev. B \textbf{64}, 195304 (2001).
\bibitem{Auston:1983} D. H. Auston, IEEE J. Quant. Electron. \textbf{QE-19}, 639 (1983).
\bibitem{Kikkawa:1998} J. M. Kikkawa and D. D. Awschalom, Phys. Rev. Lett. \textbf{80}, 4313 (1998).
\end{thebibliography}
\end{document}